# The Bitter Truth
# About Quantum Algorithms in the NISQ Era

Frank Leymann[0000-0002-9123-259X] and Johanna Barzen[0000-0001-8397-7973]

University of Stuttgart, IAAS, Universitätsstr. 38, 70569 Stuttgart, Germany
`{firstname.lastname}`@iaas.uni-stuttgart.de

**Abstract.** Implementing a quantum algorithm on a NISQ device has several challenges that arise from the fact that such devices are noisy and have limited quantum resources. Thus, various factors contributing to the depth and width as well as to the noise of an implementation of an algorithm must be understood in order to assess whether an implementation will execute successfully on a given NISQ device. In this contribution, we discuss these factors and their impact on algorithm implementations. Especially, we will cover state preparation, oracle expansion, connectivity, circuit rewriting, and readout: these factors are very often ignored when presenting an algorithm but they are crucial when implementing such an algorithm on near-term quantum computers. Our contribution will help developers in charge of realizing algorithms on such machines in (i) achieving an executable implementation, and (ii) assessing the success of their implementation on a given machine.

## 1. Introduction

It is well known that the prerequisites of a quantum algorithm must be very carefully considered when determining its applicability for a certain problem - this is nicely referred to as the algorithm's "fine-print" in [1]. In addition, when implementing and executing an algorithm on a particular near-term device (a.k.a. a NISQ machine [54]) the depth and width of this algorithm is key for assessing the potential of its successful execution on that particular machine: roughly, the *depth* of an algorithm is the number of gates to be performed sequentially (see section 5.1 for a precise definition), and its *width* is the number of qubits it actually manipulates.

Each algorithm has to be transformed in a manner that is specific to the quantum hardware it should execute on. This transformation is performed by the quantum compiler of the target machine. The compiler will both, (i) map the gate set used by the implementation of the algorithm to the gate set supported by the target machine, and (ii) it will try to optimize the width and depth of the implementation of the algorithm in dependence of the concrete hardware [8], [30]. (Note, that we assume a gate-based approach to quantum computing throughout this paper.) This already indicates that the concrete hardware a certain algorithm should be executed on has a big impact on the success of its execution. A framework for such compilers is presented in [33]. But hardware independent optimizations can be achieved too [17], [27], [65].

## 1.1. The Impact of Hardware on Quantum Algorithms

First, todays quantum computers are *noisy*, i.e. their qubits are erroneous, meaning that their actual states are not stable: the states decay over short periods of time. Similarly, the implementations of the gates used to manipulate the qubits are erroneous to, i.e. a gate does not manipulate the qubits it operates on exactly, resulting in small deviations from the expected result, and such errors propagate in course of the execution of an algorithm. The phenomenon that qubits are erroneous is referred to as decoherence, the phenomenon that gates are erroneous is referred to as gate infidelity [47], [56].

Second, todays industrial technologies of building quantum computers prohibit large numbers of qubits on a device: many qubits on a single device as well as the technology to control and connect them introduce disturbances of the qubits. Thus, todays technology has limited scalability, it is of *intermediate scale* only. Being noisy and of intermediate scale coined the term Noisy Intermediate Scale Quantum (NISQ) computers [54].

Because of noise and the limited number of qubits of NISQ machines, the depth d of a quantum algorithm as well as its width w must be controlled. In practice, the following simple formula results in a rule-of-thumb that is very helpful to assess the limits of executing a quantum algorithm on a given quantum computer:

$$d \cdot w \ll \frac{1}{\varepsilon}$$

In this formula $\varepsilon$ is the *error rate* of the quantum computer. Informally, the error rate subsumes decoherence times of qubits, precision and error frequency of gates etc; a formal and detailed discussion is given e.g. in [76]. For our purpose, the detailed definition of the error rate is not needed, an informal understanding suffice. As implied by the formula, the depth d or the width w have to be "small". For example, if an algorithm requires 50 qubits (w=50) and it should run on a quantum computer with an error rate of about $10^{-3}$ ($\varepsilon \approx 10^{-3}$), then d must be (significantly) less than 20 (d≪20), i.e. the algorithm has to complete after at most 20 sequential steps - otherwise the result would be much to imprecise.

Algorithms that require "few" qubits only can be simulated on classical computers. N qubits imply to store $2^N$ amplitudes of the state of the corresponding N qubit quantum register. As a simplistic estimation (one byte for each amplitude) $2^N$ bytes are needed to store the quantum state of N qubits. Thus, for 20 qubits $2^{20} = (2^{10})^2 \gtrsim (10^3)^2 = 10^6$ bytes are required, i.e. the corresponding quantum state will fit into the main memory of most computers, while for 50 qubits $2^{50} \gtrsim 10^{15}$ bytes are required which is hard even for large supercomputers today (although in special situations a little higher number of qubits can be simulated [10]).

Thus, quantum algorithms that are able to prove the power of quantum computers will make use of many qubits because otherwise these algorithms can be simulated on classical computers. According to the formula above, such an algorithm must have a low depth, and the algorithm is then called a *shallow* algorithm. But if an algorithm requires a high depth (a so-called *deep* algorithm) it can make use of a few qubits only and it can, thus, be simulated. This implies that quantum advantage has to be shown based on shallow algorithms, i.e. algorithms that use only "a few" layers of parallel gates (see section 5.1).

**1.2. The Consequence**

The low depth of shallow algorithms implied by the capabilities of todays NISQ machines mandate to realize *hybrid* algorithms, i.e. algorithms that are split into classical parts and shallow quantum parts. The classical parts and quantum parts of a hybrid algorithm are performed in a loop where results are optimized from iteration to iteration [20], [78]. Optimization may affect both, the quantum parts of a hybrid algorithm as well as its classical parts [26].

Figure 1 shows the general structure of a hybrid algorithm (dashed elements indicate optionality). Its quantum part is structured like any quantum algorithm: it prepares the state to be manipulated by the unitary transformation representing the algorithm proper, and the result produced by the unitary transformation is measured and passed to the classical environment. If the measured result is satisfiable (e.g. it has a certain precision) it is delivered as the final result. Otherwise the measured result is postprocessed (e.g. parameters that are used to influence the creation of the quantum state are modified), and the output of the postprocessing is passed as input to a preprocessing step that controls the state preparation accordingly.

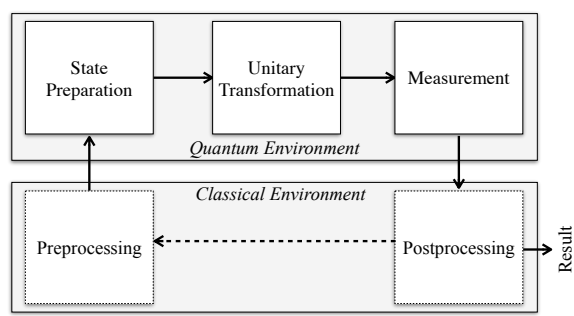

**Fig. 1**. General structure of a quantum algorithm (adapted from [39]).

This structure of a hybrid algorithm is also the generic, principle structure of most quantum algorithms: the quantum state processed by the algorithm proper (i.e. the unitary transformation) must be prepared which typically requires classical preprocessing. The output produced by measuring the result of the algorithm proper must be postprocessed, e.g. by assessing or improving its quality. In many cases the postprocessing step will not kickoff another iteration, i.e. it will not pass input to the preprocessing step (this optionality is indicated in the figure by the dashed format of the arrow between post- and preprocessing).

As a consequence, analyzing the unitary transformation, i.e. the algorithm proper is by far not enough for assessing whether or not an algorithm can be successfully executed on a NISQ device. State preparation (tightly interwoven with preprocessing) as well as measurement (tightly interwoven with postprocessing) must be considered to fully understand the requirements of an algorithm and to determine the depth and width of the circuit finally to be executed (see sections 2 and 6). Also, the algorithm proper must be analyzed, e.g. if it makes use of oracles: these subroutines contribute themselves to the depth and width of the overall circuit to be performed (see section 3). The target machine on which the circuit implementing the algorithm should run has its impact on the overall circuit to be executed because it might support a different gate set than assumed by the algorithm (see section 5) or because the target machine reveals certain hardware restrictions (see section 4).

**1.3. Contributions**

The three high-level contributions of this paper are (i) to raise awareness of developers of quantum software in practice about the fact that the unitary transformation at the core of an algorithm (which is often the focus of its potential users) is not sufficient to assess its successful executability on a certain NISQ machine, (ii) to discuss the main sources of complexity increase of an algorithm when preparing it for execution on a certain NISQ machine, and (iii) to present details about tasks that have to be performed to transform an algorithm into a circuit ready to be executed on a NISQ machine.

The need for these contributions has been triggered by our own work on using quantum algorithms in the humanities [4], [5]: We are working on a component that analyses quantum algorithms and evaluates their fit for certain NISQ machines [57], [39]. We are developing a platform [38] as part of the PlanQK project [55] that includes the before-mentioned analysis component. Also, we work on a pattern language for quantum computing to support practitioners in developing (hybrid) quantum applications: an initial version of this pattern language has been proposed in [37], and what we describe in this contribution will result in several extensions of this pattern language.

The structure of the paper is as follows: we discuss major related work in the introduction of each section. Section 2 discusses various techniques of how to prepare a quantum state needed as input for a quantum algorithm proper. The possibly deep impact on an algorithm caused by the need to implement each oracle assumed by an algorithm is subject of section 3. In section 4 the influence of the concrete topology of a quantum chip on the circuit realizing an algorithm is presented. Rewriting a circuit of an algorithm because of hardware idiosyncrasies is subject of section 5. Readout errors and the implied additional steps to correct them are sketched in section 6. All this needs additional information about the algorithm, but especially about the hardware used, its static and dynamic properties: section 7 very briefly outlines this. Section 8 gives an outlook on the impact of correcting errors of individual qubits. A conclusion and description of our ongoing and future work closes the paper.

Appendix A presents four experiments performed on several quantum computers of the IBM Quantum Experience family that show the impact of the error rates of the topologies of these machines while transpiling quantum algorithms. Three rows of the calibration matrix used to correct readout errors are determined in another experiment that is sketched in appendix B. Finally, appendix C discusses experiments we run to check the practicality of simple error correction on a NISQ machine.

## 2. Input Preparation

A quantum algorithm in general assumes input data that is required for producing a result. In order to be manipulated on a quantum computer, this input data must be represented as a quantum state. In this section we discuss several mechanisms to deliver such a state as input for a quantum algorithm, jointly referred to as *input preparation*.

In general, input preparation consists of two parts: a classical part that creates a circuit that can then be processed on a quantum computer to prepare the quantum state. The latter part is referred to as *state preparation*, while the former part is subsumed by *preprocessing*. Note, preprocessing in this case may be a manual task performed by a human (e.g. crafting a circuit) or an automatic task performed by a program (e.g. generating a circuit).

[62] discusses the number of gates required to prepare an arbitrary quantum state from classical data. Obviously, efficiency (in terms of time and space complexity) in encoding classical data into a quantum superposition state suitable to be processed by a quantum algorithm is critical [12], [44], [50]. In this section we present several such ways and their complexity. A generic mechanism to initialize arbitrary quantum states is presented in [63].

*Digital encoding* (a.k.a. binary encoding) is the representation of data as qubit strings, *analogue encoding* represents data in the amplitudes of a state. If data has to be processed by arithmetic computations, a digital encoding is preferable; mapping data into the large Hilbert space of quantum device which is often needed in machine learning algorithms, for example, prefers an analogue encoding. A brief overview on several encoding schemes like basis encoding, amplitude encoding, or product encoding can be found in [59], while [60] discusses their use in quantum machine learning.

Several algorithms require to convert a digital representation of data into an analogue representation and vice versa: [44] presents methods for mutual conversions of these representations and discusses corresponding general aspects. [12] is about optimizing the depth and width of binary encoding algorithms and present corresponding circuits. [36] abstracts some of these encodings and presents a characterization of encodings in terms of these abstractions that are (partially) robust against Pauli errors.

### 2.1. Basis Encoding

Basis encoding is primarily used when real numbers have to be arithmetically manipulated in course of a quantum algorithm. In a nutshell, such an encoding represents real numbers as binary numbers and then transforms them into a quantum state in the computational basis. [12] describes a collection of corresponding quantum circuits and methods.

More precisely, basis encoding presumes that a real number x ∈ ℝ is approximated by k decimal places and transformed into the binary representation of this approximation, i.e.

$$x \approx \sum_{i=0}^{n} b_i 2^i + \sum_{i=1}^{k} b_{-i} \cdot \frac{1}{2^i}$$

with $b_i$, $b_{-i}$ ∈ {0,1}. The sign of the real number is encoded by an additional leading binary number, e.g. "1" for "−" and "0" for "+". Thus, a real number is approximated by (n+k+2) bits, and thus prepared as a (n+k+2)-dimensional quantum state.

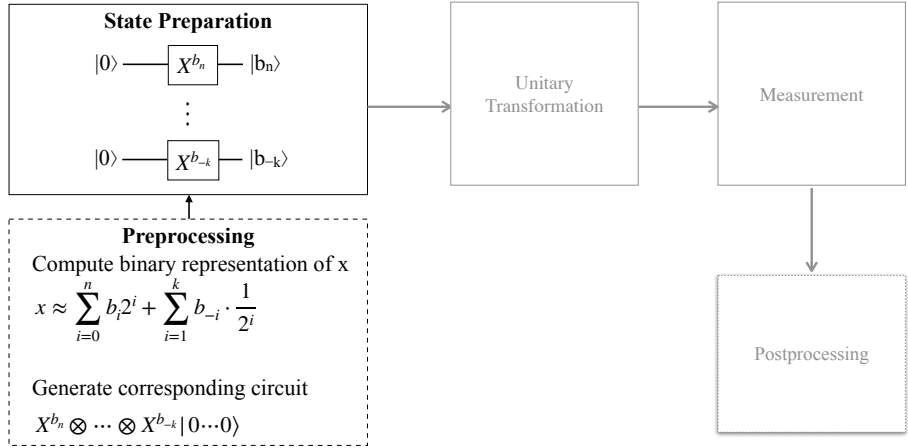

**Fig. 2**. Preparing input based on basis encoding.

Figure 2 illustrates the proceeding. In a classical preprocessing step, the binary approximation $b_s\ b_n \cdots b_0, b_{-1} \cdots b_{-k}$ of the real number is computed (where $b_s$ is the bit encoding the sign of the number). Next, the corresponding circuit is generated by applying the gate $X^{b_i}$ on qubit i. This circuit will create the quantum state representing the real number x. Then, the generated circuit is prepended to the circuit of the algorithm proper (represented by the unitary transformation in the figure) and executed on a quantum device.

The circuit can be generated in a straightforward manner as OpenQASM [13] code or Quil [66] code, for example. This code can be send to and executed on quantum computers from IBM or Rigetti, respectively. In general, other quantum instruction languages can be also used in order to generate the state preparation circuits.

Obviously, any vector x = ($x_1$,…,$x_n$) ∈ ℝⁿ can be transformed into a basis encoding by computing the basis encoding of each of its component $x_i$ and concatenating the resulting encodings. For example, $(-0.7, 0.1, 0.2)^T \in \mathbb{R}^3$ is encoded as $|x\rangle = |11011\ 01011\ 00011\rangle$.

When a data set $D = \{x^1,\ldots,x^m\} \subseteq \mathbb{R}^n$ has to be processed, the representation of D in binary encoding is the uniform superposition of the binary encoded states of the elements of D:

$$|D\rangle = \frac{1}{\sqrt{m}} \sum_{i=1}^{m} |x^i\rangle$$

An algorithm of time complexity $O(mn)$ to create such a superposition has been described in [75] (see [60] for an instructive description of this algorithm). The created superposition is called *quantum associative memory*.

The ability to access individual elements of a state representation of a data set $D=\{x^1,\ldots,x^m\} \subseteq \mathbb{R}^n$ of binary encoded states is achieved by the following so-called *quantum random access memory* (QRAM) representation:

$$\frac{1}{\sqrt{m}} \sum_{j=1}^{m} |j\rangle |x_1^j, \cdots, x_n^j\rangle$$

[24] introduces an architecture for implementing such a quantum random access memory with time complexity $O(\log m)$. Data from $\mathbb{R}^n$ can be encoded in its memory cells in $O(\sqrt{n})$. [53] suggests an implementation of QRAM that reduces the complexity of encoding to $O(\text{polylog } n)$ by assuming minimal hardware-assisted preprocessing. A flip-flop QRAM is proposed by [50] jointly with an implementing circuit which requires $O(mn)$ steps for preparing and updating states. Note, that no efficient implementation of QRAM is available as of today.

## 2.2. Amplitude Encoding

If arithmetic manipulation of data by quantum algorithms is not in the foreground, more compact representations of data are used. Especially, the large Hilbert space of a quantum device is properly exploited in such encodings.

The amplitude encoding of $x \in \mathbb{R}^n$, $\|x\| = 1$, is the quantum state $|x\rangle = \Sigma x_i |i\rangle$. This encoding requires $\lceil \log_2 n \rceil$ qubits to represent an n-dimensional data point. The state can be created by a unitary transformation $U=U_1 \otimes U_2 \otimes \ldots \otimes U_k$ where each $U_i$ is either a 1-quit gate or a CNOT, and k is of order $4^n$ for arbitrary x. [45] presents an algorithm to determine such a decomposition of an arbitrary unitary transformation.

In case x should be prepared from the $|0\rangle$ base state (i.e. $|x\rangle=U|0\rangle$), $2^n$ gates suffice [62]. This complexity can be reduced further for x that satisfy certain constraints: for example, [67] gives a polynomial algorithm for unit-length states that are specifically bounded. I.e. if exponential complexity in preparing quantum states should be avoided, only a limited set of states can be prepared.

For a non-unit length vector $x=(x_1,\ldots,x_n) \in \mathbb{R}^n \setminus \{0\}$, the amplitude encoding is

$$|x\rangle = \sum \frac{x_i}{\|x\|} |i\rangle$$

[58] presents an amplitude encoding for not necessarily unit-length vectors that especially shows how the often made assumption that n is a power of 2 can be removed by padding. For sparse vectors $x \in \mathbb{R}^n \backslash \{0\}$, [53] suggests an algorithm for amplitude encoding with improved time complexity.

### 2.3. (Tensor) Product Encoding

While the amplitude encoding requires only $\lceil \log_2 n \rceil$ qubits to encode $x \in \mathbb{R}^n$, the preparation of this state is in general exponentially expensive. The *(tensor) product encoding* discussed in the following requires n qubits to represent n-dimensional data but is in terms of complexity cheaper to prepare: it requires one rotation on each qubit. This encoding is directly useful for processing data in quantum neural networks, and is referred to in this context as *angle encoding* [60].

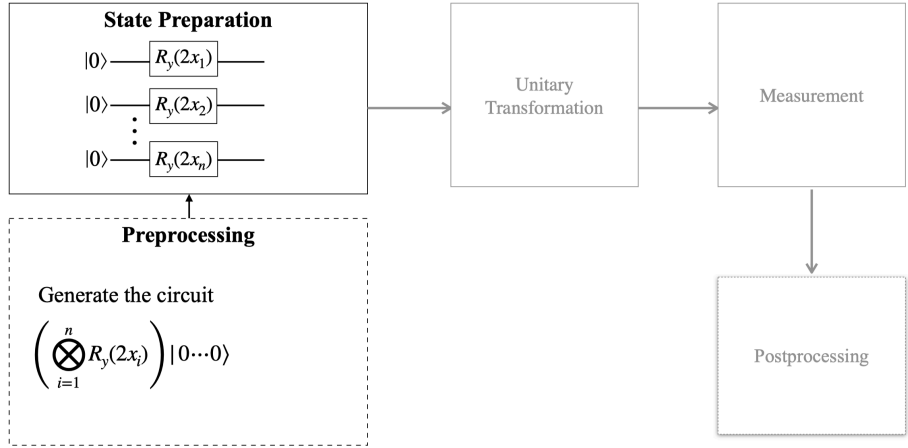

**Fig. 3**. Preparing input based on product encoding.

For $x=(x_1,\ldots,x_n) \in \mathbb{R}^n$ the tensor product encoding represents each component $x_i$ as $|x_i\rangle = \cos x_i \cdot |0\rangle + \sin x_i \cdot |1\rangle$. The complete vector as represented as

$$|x\rangle = \begin{pmatrix} \cos x_1 \\ \sin x_1 \end{pmatrix} \otimes \cdots \otimes \begin{pmatrix} \cos x_n \\ \sin x_n \end{pmatrix}$$

With the unitary rotation operation

$$R_y(2x) = \begin{pmatrix} \cos x & -\sin x \\ \sin x & \cos x \end{pmatrix}$$

it is $R_y(2x_i)|0\rangle = \cos x_i \cdot |0\rangle + \sin x_i \cdot |1\rangle = |x_i\rangle$. Thus,

$$\left( \bigotimes_{i=1}^{n} R_y(2x_i) \right) |0\cdots 0\rangle = \begin{pmatrix} \cos x_1 \\ \sin x_1 \end{pmatrix} \otimes \cdots \otimes \begin{pmatrix} \cos x_n \\ \sin x_n \end{pmatrix} = |x\rangle$$

The preprocessing step can generate the corresponding circuit such that the state preparation step will prepare the quantum representation of x (see Figure 3). As before, the circuit can be generated as OpenQASM or Quil code, for example, prepended to the algorithm proper, send to a quantum computer which is then executed there.

## 2.4. Schmidt Decomposition

Let V, W be two Hilbert spaces, dim V=n $\geq$ m=dim W. Then, for any x $\in$ V⊗W there exist orthonormal sets $\{u_1,\ldots,u_m\} \subseteq$ V, $\{v_1,\ldots,v_m\} \subseteq$ W and $\alpha_1,\ldots\alpha_m \in \mathbb{R}_{\geq 0}$ with $\Sigma\alpha_i=1$ such that:

$$x = \sum_{i=1}^{m} \alpha_i \cdot u_i \otimes v_i$$

The numbers $\alpha_1,\ldots\alpha_m$ are uniquely determined by x and are called the *Schmidt coefficients* of x, the corresponding representation of x is called its *Schmidt decomposition*. For x $\neq$ 0 there exists K $\in \mathbb{N}$ (which is uniquely determined by x, the so-called *Schmidt rank* of x) such that $\alpha_1 \geq \alpha_2 \geq \ldots \geq \alpha_K > \alpha_{K+1} = \ldots = \alpha_m = 0$. Note that x is entangled if and only if K>1.

The Schmidt decomposition of x can be computed by means of the singular value decomposition of a certain matrix. [25] introduced an algorithm to compute the singular value decomposition of a matrix (see [40], [49] for contemporary presentations). Based on this, the Schmidt decomposition is computed [2]: first, orthonormal bases $\{e_i\}$ and $\{f_j\}$ for V and W are chosen and x is represented as linear combination in the basis $\{e_i \otimes f_j\}$ of V⊗W:

$$x = \sum_{i,j} \beta_{ij} \cdot e_i \otimes f_j$$

Next, the singular value decomposition of M:=$(\beta_{ij})$ is computed, i.e.

$$M = (U_1 \ U_2) \begin{pmatrix} A \\ 0 \end{pmatrix} V^*$$

with unitary U=$(U_1 \ U_2)$, unitary V, and a positive-semidefinite diagonal matrix A. The column vectors of $U_1$ are then the vectors $\{u_1,\ldots,u_m\}$, the column vectors of V are the vectors $\{v_1,\ldots,v_m\}$, and the diagonal elements of A build the set $\{\alpha_1,\ldots\alpha_m\}$.

In practice, sometimes a quantum register is split into two parts and the entanglement of the state between these two parts must be controlled; state preparation based on Schmidt decomposition can be used in this case [15]. Figure 4 shows the circuit of the algorithm to initialize a quantum register in state x$\in$V⊗W. There are differences to consider for quantum registers with even or odd numbers of qubits: see [52] for details.

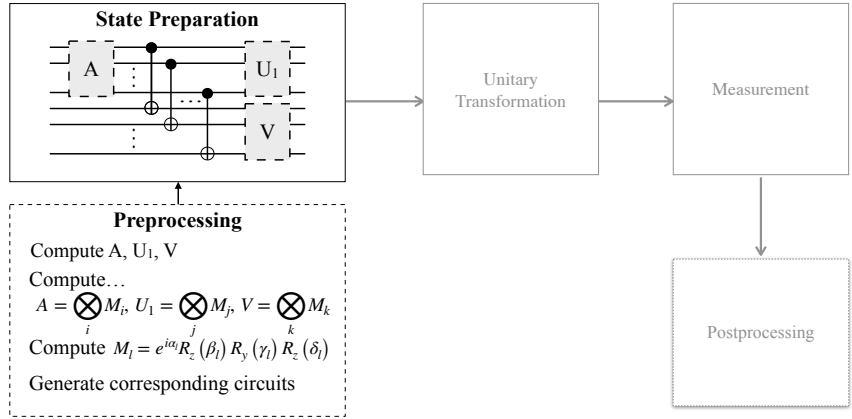

**Fig. 4**. Preparing input based on Schmidt decomposition.

In the upper left part of the figure, A, U1, V are rendered with dashed borders and grey shaded: in order to be executable on a quantum device, these unitary operators have to be represented as a composition of a universal set of 1-qubit and 2-qubit gates, i.e. as subroutines themselves: $A=\otimes M_i$, $U1=\otimes M_j$ and $V=\otimes M_k$.

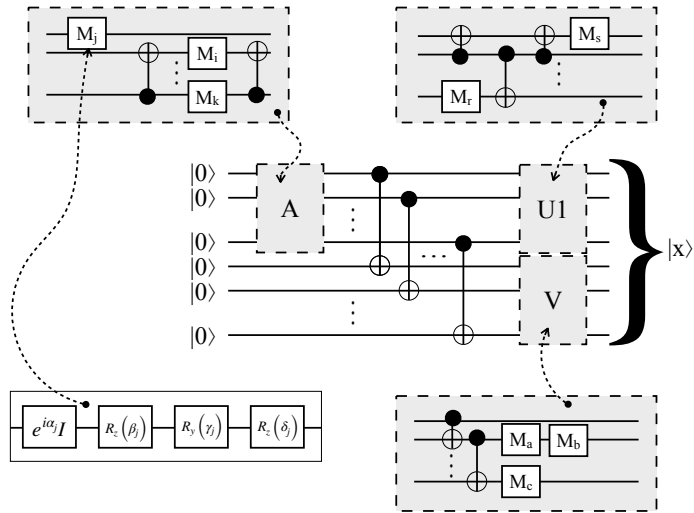

**Fig. 5**. Representing the unitary operators of the Schmidt decomposition as subroutines.

Figure 5 indicates this by showing that A, U1 and V are rewritten as separate schematic circuits using 1-qubit operators and CNOTs (see [47], [56] for a general treatment of representing multi-qubit gates via 1-qubit and 2-qubit gates, and [45] for a specific algorithm to determine such a representation). Furthermore, each of the 1-qubit operation $M_l$ is represented as a so-called Z-Y decomposition, i.e. it is computed as $M_l = e^{i\alpha_l} R_z(\beta_l) R_y(\gamma_l) R_z(\delta_l)$; Figure 5 depicts the Z-Y decomposition of $M_j$. Overall, this should give an impression of the complexity of the overall circuit implementing the state preparation based on the Schmidt decomposition of the data element.

### 2.5. Conclusion on Input Preparation

In order to prepare input for a quantum algorithm as a quantum state, a quantum circuit has to be performed that prepares the corresponding state. This circuit can be generated in a classical preprocessing step - see Figure 6. The generated circuit is prepended to the circuit of the algorithm proper, send to a quantum device, and executed. The generated circuit is to be represented in the quantum instruction language of the target device.

Thus, in order to evaluate the executability of a given algorithm on a particular device, the effort and complexity in terms of additional gates and qubits required to prepare the input of the algorithm proper must be considered.

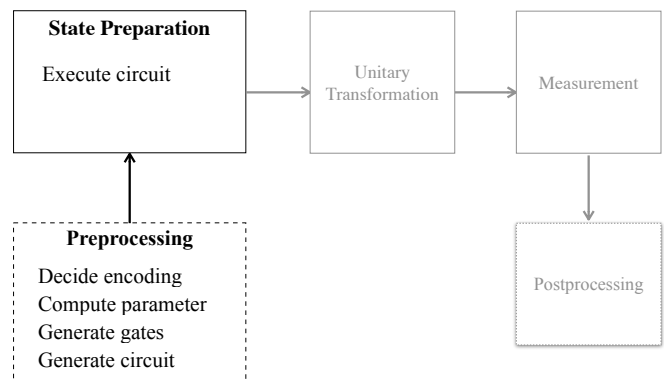

**Fig. 6**. Steps to prepare classical input for manipulation by a quantum algorithm.

## 3. Oracle Expansion

Many oracles (or black box functions in general) like the one used in the Shor algorithm for factorization, for example, perform arithmetic operations: e.g. [74], [23], or [18] provide details about implementations of such operations. Based on this, [6] gives an extensive description and analysis of modular exponentiation, which is an integral part of the Shor algorithm. [51] and [73], for example, present optimized implementations of arithmetic circuits in this context. [3] discusses the problem of oracle construction in the area of quantum clustering algorithms. Several sample oracles for the Grover algorithm are presented in [21].

The complexity of a quantum algorithm is often specified in terms of query complexity; this complexity measure considers the number of times the oracles included in the algorithms are invoked (i.e. "queried") [47]. But each oracle is implemented by a quantum algorithm itself. Thus, when an oracle is used in an algorithm it contributes (often significantly) to the depth and width of the corresponding algorithm.

Consequently, expanding all oracles of an algorithm is critical in order to assess its suitability to be executed successfully on a certain NISQ machine. Effectively, expanding oracles contributes to determine the time complexity of an algorithm in contrast to its query complexity. And time complexity (i.e. depth) is key to assess the suitability of a quantum algorithm for a NISQ machine; the same is true for space complexity (i.e. width) of a quantum algorithm.

### 3.1. Expanding the Oracles of the Shor Algorithm

Figure 7 gives a circuit for adding two numbers x and y prepared in basis encoding. The circuit acts on the register $|x\rangle \otimes |y\rangle$ that contains both summands. After executing the circuit, the first part of the register is left unchanged, while the second part contains the sum $|x+y\rangle$. Note, that the circuit makes use of the quantum Fourier transformation (QFT) and its inverse; $R_k$ is the rotation gate

$$R_k = \begin{pmatrix} 1 & 0 \\ 0 & e^{2\pi i/2^k} \end{pmatrix}$$

i.e., $R_1$ is the Pauli Z gate. It is interesting to observe that the addition circuit is similar to the circuit of the quantum Fourier transformation itself.

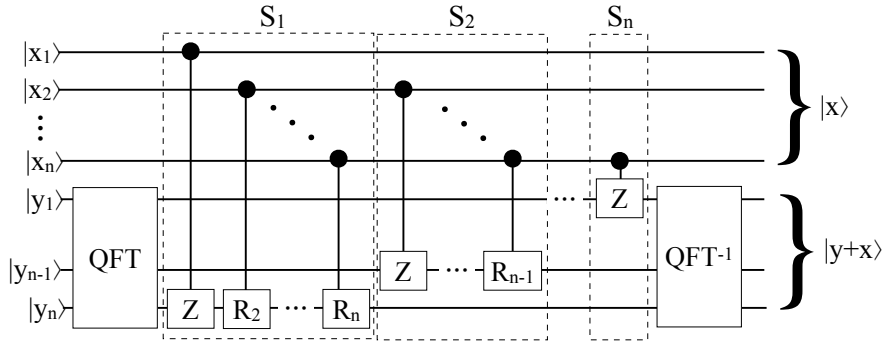

**Fig. 7**. Quantum circuit for addition (adapted from [18]).

Thus, the algorithm for adding two numbers contains other black box functions, namely QFT and QFT$^{-1}$ which have to be expanded too: oracle expansion has to happen recursively. The oracle $U_f$ of the Shor algorithm (depicted in the upper part of Figure 8) performs modular exponentiation which requires multiplication (in addition to the circuits shown in the figure) etc., i.e. the expansion of the Shor oracle is much more complex than the summation circuit shown.

Fig. 8 gives an impression of the circuit of the Shor algorithm after all black box functions are recursively expanded. The circuit at the top of the figure shows the Shor algorithm as it is given in text books: it uses two black box functions, namely the oracle $U_f$ which computes $f(x) = a^x$ mod $n$, and the quantum Fourier transformation QFT. The latter is used as a black box function to prepare the computation of the period of f via continued fractions in a classical postprocessing routine (not shown in the figure). The circuit that has to be substituted for the QFT black box function is depicted on the right side of the middle row in the figure. The left side of this row indicates the circuit that has to substitute the oracle $U_f$: this circuit consists of additions, multiplication (indicated by the ellipsis) etc. out of which just one addition circuit is depicted. The circuit of additions makes use of both, the quantum Fourier transformation QFT as well as its inverse $QFT^{-1}$, i.e. these black fox functions are to be substituted by the circuits shown in the bottom row of the figure. Together, the "lightweight version" of the Shor algorithm as given in text books becomes quite "heavyweight" by all these substitutions.

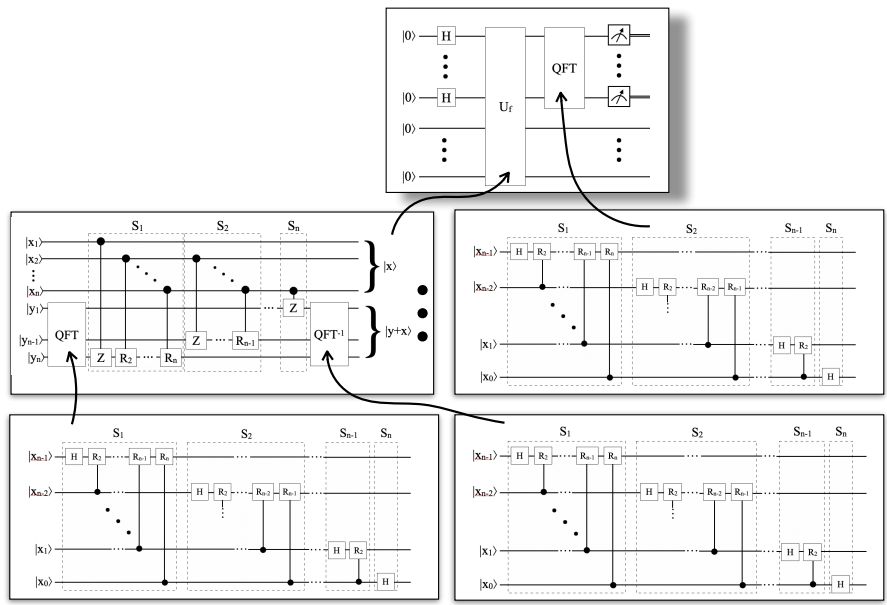

**Fig. 8**. (Partial) Expansion of black box functions of the Shor algorithm.

### 3.2. Conclusion on Oracle Expansion

A quantum algorithm that makes use of oracles (i.e. black box functions in general) must be modified by expanding these functions as quantum circuits themselves and substituting these functions by the corresponding circuits. This expansion must take place recursively, i.e. if the substituting circuits contain again black box functions they have to be substituted by further circuits. Obviously, this will often result in a significant increase of the depth of the original quantum algorithm.

In general, circuits realizing black box functions require or make also use of, respectively, ancilla qubits (this has not been discussed in the example above). Thus, expanding black box functions as quantum circuits will often increase the width, i.e. the number of qubits required by the original algorithm too.

Consequently, in order to evaluate the executability of a given algorithm on a particular device, the increase in depth and width resulting from black box expansion must be considered.

## 4. Considering Connectivity

The unitary transformations of a quantum algorithm are resolved into operations that act on a single qubit or on two qubits at once (considering CNOT suffice) [47], [56]. For actually performing 2-qubit operations the manipulated qubits must be directly, physically connected. This situation is formalized as a connected graph $G=(N,E)$ the nodes N of which represent the qubits $\{q_i|1 \leq i \leq n\}$ of the quantum computer and the edges E of which represent the direct, physical connections $\{\{q,q'\}| q,q' \in N\}$ that exist between two qubits q and q'. The graph is called the *coupling graph* or the *topology graph* of the quantum computer (e.g. [64]).

A sample topology graph G is depicted in Figure 9: the graph $G=(N,E)$ has qubits $q_1,...,q_8$ as node set $N=\{q_1,...,q_8\}$, and the edge set results from their connections $E=\{\{q_1,q_2\}, \{q_2,q_3\},..., \{q_7,q_8\}, \{q_8,q_1\}\}$. The graph G has a circular structure, and the corresponding quantum computer is said to have circular topology.

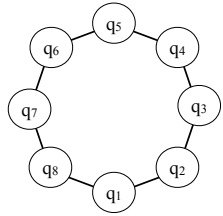

**Fig. 9**. Sample topology graph.

Like in this example, typically, not each qubit is connected to all other qubits, i.e. the graph G is not a complete graph. Thus, qubits affected by a 2-qubit operation must be moved within the graph via a sequence of so-called SWAP gates until they become adjacent (see sections 4.1 and 4.2).

Thus, the lack of connectivity introduces additional gates in form of SWAPs or it requires an appropriate mapping of the qubits of an algorithm to the physical qubits of the machine running the algorithm (see section 4.3). These additional SWAPs gates negatively impact the total error rate of the algorithm and increases its depth. Consequently, this number of SWAPs must be minimized.

[31] investigates the sensitivity of key algorithms like quantum Fourier transformation to a set of topologies, for example, linear topology, ladder topology or all-to-all topology. Implementations of these algorithms with the particular topologies in mind are presented and analyzed.

[28] presents an approach for emulating fully connected qubits on NISQ machines with limited connectivity while minimizing the impact on an algorithm's performance due to required SWAPs.

[32] considers the topology graph of a particular device to rewrite a circuit into an equivalent circuit with the minimum number of CNOT gates required to swap qubits according to a device's topology. Focussing on minimizing the number of CNOTs is justified by the fact that 2-qubit gates show a significantly higher error rate that 1-qubit gates.

This is further complicated by different error rates of individual qubits (variation-aware qubit allocation problem [22]), as well as different error rates of the connections between the qubits (qubit movement and quit allocation problem [71]).

### 4.1. The Need for SWAPs and Implications

When two qubits are subject to a 2-qubit operation these qubits must be directly, physically connected. Since in practice topology graphs are not complete graphs (i.e. not every two distinct nodes are directly connected) the qubits must be exchanged along the edges of the graph so that the two qubits to be manipulated become neighbors: this requires a sequence of SWAP operations, where each such SWAP gate switches two neighbor qubits. Thus, 2-qubit operations typically imply to inject additional gates into the circuit implementing an algorithm.

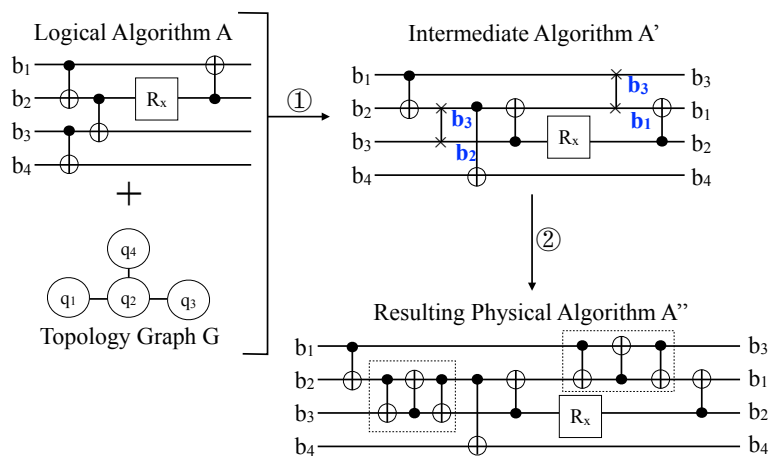

**Fig. 10**. Rewriting an algorithm based on the connectivity of a device (adapted from [32]).

Figure 10 gives a sample of such a situation. The given algorithm A is independent of any particular quantum computer (i.e. it is called a "logical" algorithm), and it should be executed on a machine with the topology graph G. The qubits manipulated by A are denoted by $b_1,\ldots,b_4$ and they are initially mapped to the (physical) qubits $q_1,\ldots,q_4$ of the machine. The algorithm A requires to perform a CNOT on $b_3$ and $b_4$ which have been initially mapped to on $q_3$ and $q_4$. Since $q_3$ and $q_4$ are not connected according to G, $b_2$ and $b_3$ are swapped, allocating $b_3$ to $q_2$ and $b_2$ to $q_3$. i.e. $b_2$ and $b_3$ exchange their position (indicated by the names in blue after the SWAP in the intermediate algorithm A'). Because $q_2$ and $q_4$ are connected in G, a CNOT on the qubits at position $q_2$ (which is now $b_3$) and the qubit at position $q_4$ (which is still $b_4$) can now be performed. Thus, CNOT($b_3$,$b_4$) in algorithm A is substituted by SWAP($b_2$,$b_3$) followed by CNOT($b_2$,$b_4$) in algorithm A'.

Note that this increased the depth of the algorithm: while CNOT($b_3$,$b_4$) and CNOT($b_1$,$b_2$) is performed in parallel in algorithm A, CNOT($b_1$,$b_2$) and SWAP($b_2$,$b_3$) and CNOT($b_2$,$b_4$) must be performed sequentially in algorithm A': an increase of the depth by 2.

Next, CNOT($b_2$,$b_3$) in A must be rewritten as CNOT($b_3$,$b_2$) in A' because $b_2$ and $b_3$ exchanged their positions. The operation $R_x(b_2)$ in A must also be rewritten: it becomes $R_x(b_3)$ in A' because of the exchange of $b_2$ and $b_3$. Algorithm A contains as last operation CNOT($b_2$,$b_1$) but the $b_2$-position is now occupied by $b_3$: a SWAP($b_1$,$b_2$) moves the actual $b_3$ (until now at position $b_2$) to position $b_1$, and the actual $b_1$ is moved to position $b_2$. After that, CNOT($b_3$,$b_2$) in A' controls the negation of $b_1$ (now at position $b_2$) by $b_2$ (now at position $b_3$). Thus, CNOT($b_2$,$b_1$) in A is substituted by the sequence SWAP($b_1$,$b_2$) and CNOT($b_3$,$b_2$) in A', with another increase of the depth of the algorithm. Thus, the topology of a quantum computer forces a rewrite of a logical algorithm which implies an increase of its depth (and a permutation of the qubits).

Furthermore, SWAP operations are typically not directly supported by quantum computers. Instead, a SWAP(a,b) operation is substituted by the equivalent sequence CNOT(a,b) and CNOT(b,a) and CNOT(a,b) (e.g. [47]). That means that each SWAP operation of the intermediate algorithm A' must be substitute by a corresponding sequence of CNOT operations resulting in the "physical" (i.e. hardware-specific adapted) algorithm A'' in Figure 10. This further increases the number of gates to be performed and the depth of the circuit implementing a logical algorithm.

### 4.2. Qubit Movement

The connections between two qubits are subject to errors too, and these errors impact the error rate of the 2-qubit operations having the connected two qubits as parameters [71]. Hence, the error rate of the circuit of an implementation of a quantum algorithm is affected by the error rates of these connections. Because of this, the connectivity graph is sometimes considered as a weighted graph where the weights of the edges are the success rates of the individual connections, i.e. the edge set E is a set of pairs (\{$q_i$,$q_j$\},$s_{ij}$) and $s_{ij}$ is the success rate of the connection between qubits $q_i$ and $q_j$. A topology graph with the weights associated with the edges is depicted in Figure 11.

The different success rates of the connections imply that minimizing the number of SWAPs is not always the right metric to determine how to move qubits that have to be jointly manipulated by a 2-qubit operation. For example, on a machine with the topology shown in Figure 11 qubits $q_1$ and $q_3$ should be manipulated by a 2-qubit operation $\Omega$. Since $q_1$ and $q_3$ are not directly connected, they have to be moved to adjacent positions. Swapping $q_3 \rightarrow q_2$, followed by $\Omega(q_1,q_2)$ has success rate 0.3×0.5=0.15. But the sequence of swaps $q_3 \rightarrow q_4 \rightarrow q_5 \rightarrow q_6 \rightarrow q_7 \rightarrow q_8$ followed by $\Omega(q_1,q_8)$ has success rate 0.8×0.8×0.9×0.9×0.7×0.9 = 0.33. In this case, a sequence of 5 SWAPs followed by Ω has a much higher success rate than the single SWAP followed by Ω.

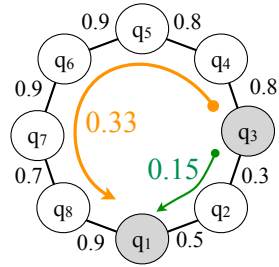

**Fig. 11**. Sample topology graph with success rates of 2-qubit operations as weights.

The success rate of a connection is influenced by several technological factors and change over time [71]. Thus, the success rates have to be determined regularly, i.e. the weighted topology graph is time dependent (see also appendix A).

**4.3. Initial Qubit Allocation**

The initial allocation of the qubits used by a quantum algorithm influences the number of SWAPs required by the circuit implementing the algorithm (e.g. [48], [64], [71]). Assigning the qubits used within the circuit to the qubits available on the target machine that should execute the circuit defines a subgraph of the topology graph of this machine. A naive approach would select any connected subgraph of the topology graph that will minimize the number of SWAPs. But in fact, different subgraphs result in different reliabilities of performing the SWAPs because of the different success rates of the connections represented by the weighted edges of the topology graph. Thus, the reliability of the implementing circuit can be improved by selecting a subgraph with maximum weights of the edges of the subgraph spanned by the initial allocation of qubits (more precisely, the product of the weights of the edges becomes the weight of the subgraph). Note, that the initial allocation will change in course of the execution of the program due to SWAPs actually performed.

Figure 12 shows an OpenQASM [13] program and the (weighted) topology graph of the quantum machine it should be executed on. The program uses three qubits Q0, Q1 and Q2, and it performs the two 2-qubit operations CNOT(Q0,Q1) and CNOT(Q2,Q1) (in OpenQASM the syntax of CNOT is cx). The table on the right side of the figure lists the various possible initial mappings of Q0, Q1 and Q2 to the set of qubits $q_1$, $q_2$,…,$q_8$ of the machine. For each such possible mapping the success rates of the edges of the connected subgraph defined by each mapping are multiplied and assigned as weight to the mapping. Based on this table, it turns out that the initial assignment Q0↦$q_5$, Q1↦$q_6$ and Q2↦$q_7$, ist the optimal initial allocation of qubits.

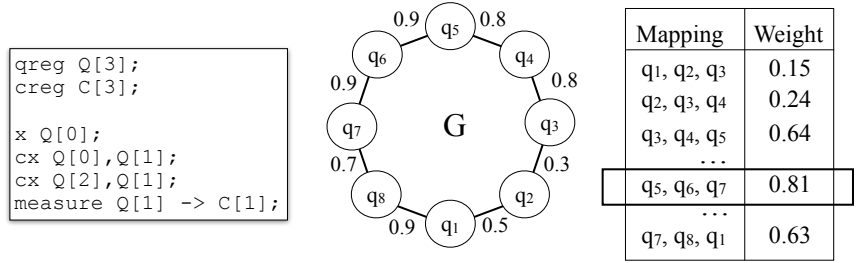

**Fig. 12**. Initial qubit allocation minimizes number of SWAPs required (adapted from [71]).

The initial qubit allocation is also influenced by the error rates of 1-qubit operations acting on each particular qubit ([48], [64], [71]). For this purpose the topology graph becomes a colored graph in which the success rate of certain 1-qubit operations is associated with each node. Since 1-qubit operations are about one order of magnitude less erroneous than CNOT, the approaches to initial qubit allocations mentioned here ignore errors of 1-qubit operations. Executing quantum programs on machines available via IBM Quantum Experience instead show that the corresponding compiler considers these 1-qubit errors, and that these error rates change over time too (see appendix A).

### 4.4. Conclusion on Connectivity

The connectivity of a quantum computer as represented by its topology graph has impact on the depth of a circuit implementing a quantum algorithm. This is because the connectivity typically requires a rewrite of the circuit to enable 2-qubit gates of the circuit. Such a rewrite results in the injection of sequences of CNOT operations that increases the depth of the circuit. The different success rates of the connections of two qubits of a target machine further impacts the increase of depth of the circuit.

Hence, in order to evaluate the executability of a given algorithm on a particular quantum computer, the increase in depth as enforced by the topology graph of the machine must be considered. For example, this increase in depth can be understood by analyzing the output of the compiler (or transpiler) of the corresponding target machine (see section 5.3).

## 5. Circuit Rewriting

Before executing a circuit on a particular quantum computer, the corresponding circuit is typically rewritten. For this purpose, the corresponding environments include a compiler that rewrites the circuit such that it can be performed on the target machine [39]. Since such a rewrite is highly dependent of the concrete machine, such a compiler is sometimes referred to as a cross-compiler or transpiler.

The foremost reason for such a rewrite is to map the gates used by a circuit to the gates that are physically supported, i.e. directly implemented, by the target quantum computer. Also, the time the qubits of the circuit are used in excited states is minimized to reduce noise and error rates. Another important reason for rewrite is to strive to reduce the depth and width of the circuit. Finally, rewrite considers the connectivity of the target device. Altogether, rewriting optimizes a circuit for the execution on a particular quantum computer.

### 5.1. Reducing the Depth of a Circuit

A *level* of a circuit is a collection of its gates that can be performed in parallel. All gates of a certain level are executed within the same time unit. The gates of the succeeding level are performed once all gates of the preceding level finished. In general, the higher the degree of parallelism the shorter is the duration of the execution of the overall algorithm. A shorter duration corresponds to a reduction of the number of errors due to decoherence. This increases the robustness and precision of the implementation of the algorithm. Thus, having a minimum number of levels of a circuit — its so-called *depth* — is a key aspect of a circuit.

It is important to note that a circuit specifies the data flow between its gates, not their control flow. Thus, gates of succeeding levels that manipulate different qubits can be shifted to belong to the same level. Iteratively, gates at different levels that manipulate different qubits can be shifted into the same level. Within a circuit time is passing from left to right, hence gates are to be moved to the leftmost level possible because this increases the probability that the qubits they manipulate have not yet been subject to decoherence. This way reliability and precision of the execution of a circuit is improved. Levels that are left empty after shifting all gates are deleted. Finally, the depth of the circuit has been reduced.

In Figure 13 the circuit on the right has been originally specified in four levels. Level 2, for example, consists of gates G21 and G22. Gates G11 and G21 are manipulating different qubits, thus G21 can be shifted from level L2 to level L1; the same can be done with gate G22. This leaves level L2 empty, so G32 is shifted to L2, and, finally, G41 and G42 can be shifted to L2 too. After that levels L3 and L4 are empty, therefor they can be deleted: the remaining circuit has two levels, i.e. it is of depth 2. As a result, the depth of the original circuit has been reduced by "shifting gates to the left as far as possible".

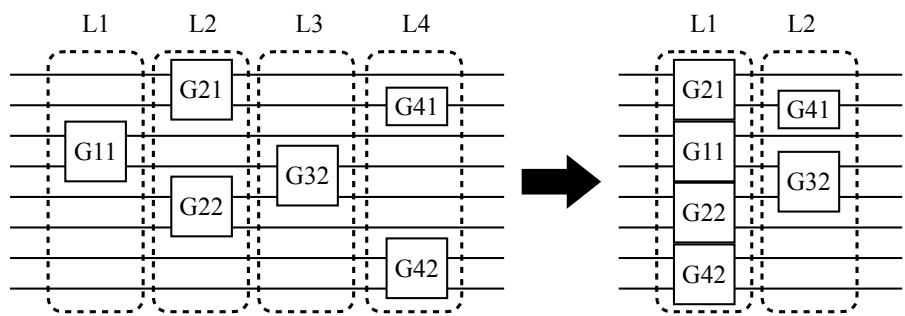

**Fig. 13**. Reducing the depth of a circuit.

### 5.2. Circuit Optimization

Reducing the number of gates of a circuit obviously decreases the number of gate errors, contributing to its robustness and precision. [65] describes an optimizing compiler that is hardware independent and is geared towards NISQ devices. The intermediate representation of the compiler is an abstract gate-based circuit model, thus it is independent of the various frontend programming languages used. The compiler acts in two phases: (i) an optimization phase reduces the size of the circuit by reducing the number of 2-qubit gates as well as the depth of all 2-qubit gates in a hardware independent manner, and (ii) a hardware-specific phase that maps the circuit to a certain device including a gate mapping and a qubit mapping. This reduces the transpilation effort of the target environment. [43] uses a generalization of rewriting rules called templates to first reduce the number of gates of a circuit and reduce its depth afterwards.

[77] is rewriting QASM programs to reduce the time a qubit is used, especially in excited state. This diminishes the overall error in quantum circuits by decreasing the probability that the corresponding qubit state is subject to decoherence.

### 5.3. Transpilation

The principle phases of transforming a quantum algorithm into a circuit using instructions that are physically implemented on a specific quantum device has been described in [70].

The transpiler especially solves the (variation-aware) qubit allocation problem as well as the (variation-aware) qubit movement problem (see sections 4.2 and 4.3). In doing so, the transpiler will change gates specified in the algorithm and will add new gates to it. This might significantly impact the depth of the original algorithm which directly influences the appropriateness of the target device of the transpiler.

In the following, we sketch a few experiments we run on different devices of the quantum computers offered by IBM Quantum Experience.

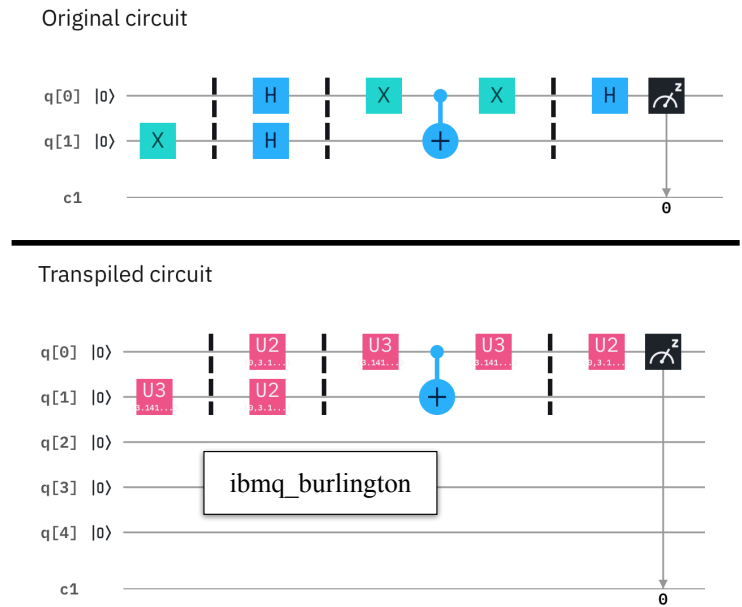

**Fig. 14**. Transpilation substitutes gates by physically implemented gates.

Figure 14 shows an experiment we performed on the ibmq_burlington machine. The original circuit is shown in the upper half of the figure; it is an implementation of the Deutsch algorithm [47] with the (simple) oracle expanded. The result produced by the transpiler is depicted in the lower half of the figure: all 1-qubit gates used in the original circuit have been replaced by the rotation gates which are directly implemented by the hardware of the device. Otherwise, the data flow of the algorithms are identical.

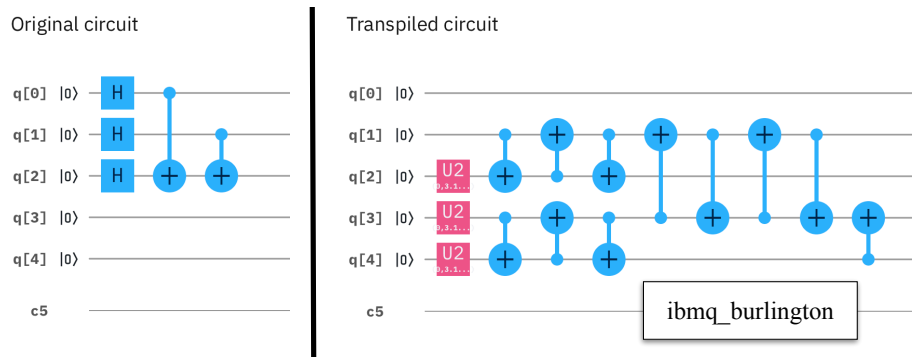

**Fig. 15**. Transpilation may increase the depth of a circuit.

The next experiment (see Figure 15) was executed on the same machine. As before, all 1-qubit gates of the original algorithm are substituted (i.e. the Hadamard H gates are implemented by U2(0,$\pi$) gates). Furthermore, the initial qubit allocation assigned different physical qubits to the qubit variables of the original algorithm. Several SWAPs have been consequently added, which increased the depth of the algorithm from 3 of the original algorithm to 9 of the transpiled algorithm.

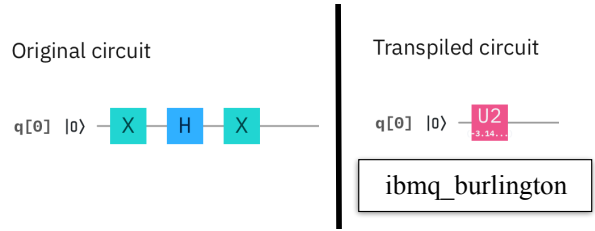

**Fig. 16**. Transpilation may decrease the depth of a circuit.

Figure 16 shows another experiment we performed again on the ibmq_burlington machine. The original circuit has three sequential gates and these three gates are substituted by a single U2 gate (with proper angles). Thus, while the original circuit has a depth of 3, the transpiled circuit has a depth on 1, i.e. the depth has been reduced.

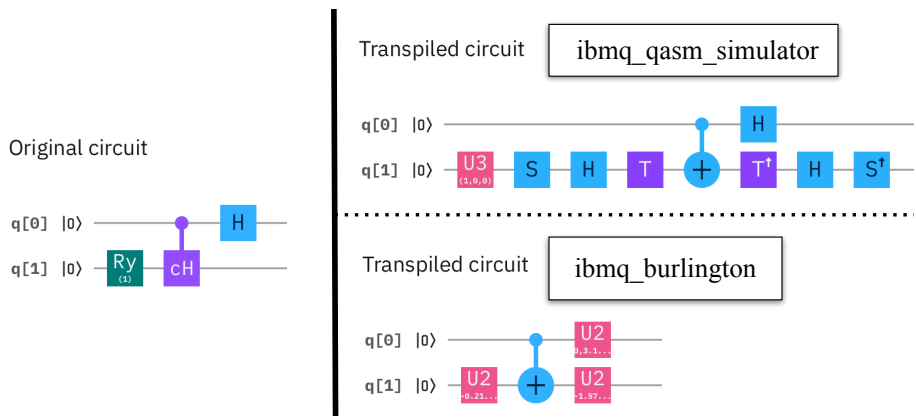

**Fig. 17**. Transpilation is device dependent.

The final experiment on transpilation has been conducted to show the dependencies of the result of the transpilation process on the target hardware: the original circuit (shown on the left side of Figure 17) has been transpiled to two different devices. The result is depicted on the right side of the figure: the upper part represents the transpilation to ibmq_qasm_simulator (a simulator delivered by IBM Quantum Experience), the lower part the transpilation to ibmq_burlington. It can be seen, that both results are quite different: while the transpilation to ibmq_burlington did not change the depth of the original circuit, it added one more 1-qubit gate; the depth of the transpilation to ibmq_qasm_simulator increased the depth of the original circuit from 3 to 8 and introduced sequences of quite different gates than in the original circuit.

### 5.4. Conclusion on Circuit Rewriting

The circuit implementing an algorithm can often be optimized by reducing its depth. This is achieved by increasing the degree of parallelism in which the gates of the circuit are executed without sacrificing the specified data flow of the algorithm ("shifting gates to the left as far as possible").

Typically, an algorithm is specified in a hardware independent manner. But in order to be executed on a particular quantum computer it must be transformed to fit the physical realities of the target machine. This transpilation process has again impact on the depth of a corresponding circuit. Thus, the transpiled circuit has to be inspected to assess the suitability of the target machine for successfully executing a given algorithm.

## 6.  Correcting Readout Errors

Measuring a qubit takes significantly longer than unitary operations on qubits. Thus, during measurement, the qubits being measured may change their states because of decoherence introducing so-called *readout errors*.

[46] describes several unfolding methods (see section 6.1) to correct readout errors and applies the "matrix inversion" method to a device from the pool of the IBM Quantum Experience machines. Unfolding means to adapt the measured result distribution to a most probable true result distribution.

[41] proposes a method that uses postprocessing based on quantum detector tomography that reconstructs the measurement device. [42] uses machine learning techniques to improve the assignment fidelity

$$F = 1 - \frac{1}{2}\left(\mathbb{P}(0\,|\,1) + \mathbb{P}(1\,|\,0)\right)$$

of single qubit measurements up to 2.4% in their experiments ($\mathbb{P}(m|s)$ denotes the probability that the outcome m is measured but the state $|s\rangle$ was prepared). The following section presents a straightforward method with acceptable results for dealing with readout errors on small NISQ machines.

### 6.1.  Unfolding

When data is measured, the measuring device may introduce errors, i.e. the measured value does not exactly reflect the true state of the system measured. Repeating such a measurement results in a distribution of measured values which deviates from the true distribution of values that would have resulted if the measurement device would be error free. *Unfolding* is a method to determine the true, undisturbed distribution from the measured, disturbed distribution.

Reading out data from a quantum computer results in a bit string $(b_0,\ldots,b_{n-1})$, which is represented in decimal encoding as a natural number $k \in \{0,\ldots,2^n-1\}$.

Running the same computation several times will result in a distribution where the value k is measured count(k) ∈ ℕ times. Thus, the result is a vector of natural numbers whose i-th component is count(i), i.e. the frequency of how often i has been measured. Let t be the vector corresponding to the true, undisturbed distribution and m be the vector of the measured, disturbed distribution. Figure 18 depicts these two distributions as histograms where the white bins represent the true distribution while the gray bins represent the measured distribution.

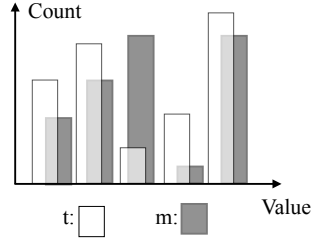

**Fig. 18**. True and measured readout distributions.

Several methods can be used to approximate the true distribution from the measured distribution: [9] discusses several such unfolding methods and compares them. As a first approach, the so-called matrix inversion method is applicable in our context. This method assumes that t and m are related by a *calibration matrix* C (a.k.a. response matrix a.k.a. correction matrix) such that m = C·t. The coefficient $C_{ij}$ of the matrix C is the probability of measuring the value i under the condition that the true value is j:

$$C_{ij} = \mathbb{P}(\text{measured value is i} \mid \text{true value is j})$$

Hence, the assumption m = C·t gives

$$m_i = \sum_k C_{ik} t_k = \sum_k \mathbb{P}(i \mid k)\, t_k$$

and with $m_i \approx \mathbb{P}(i)$ and $t_k \approx \mathbb{P}(k)$ the assumption becomes the Law of Total Probability, i.e. the assumption becomes justified:

$$\mathbb{P}(i) = \sum_k \mathbb{P}(i \mid k)\, \mathbb{P}(k)$$

The further assumption of the matrix inversion method is that the calibration matrix C is regular, i.e. $C^{-1}$ exists. Then,

$$t = C^{-1} \cdot m$$

[46] explores the matrix inversion method in depth on IBM Quantum Experience devices and compares it by means of experiments with other unfolding methods. The finding is that for sufficiently small readout errors the results are all acceptable, although matrix inversion is not the best method.

Obviously, this method is only practical for small numbers n of qubits because $2^n$ measurements are required. However, this is consistent with our assumption that we are dealing with NISQ devices which have low number of qubits. But even determining the $2^n \times 2^n$ calibration matrix for a device with 20 qubits seems to be out of scope in practice (because $2^{20} = (2^{10})^2 \approx (10^3)^2 = 10^6$).

Determining the calibration matrix (see section 6.2) assumes that the state to be measured has been accurately prepared. Typically, this is not the case, i.e. determining the calibration matrix is also strained with state preparation errors which results in the name *SPAM errors* (state preparation and measurements errors) for this combined class of errors.

While in general determining the calibration matrix for an n qubit device requires an exponential number of measurements, [69] describes a method with a quadratic number of measurements but the reliability of the matrix derived is observed to be accurate for chains of qubits mostly.

### 6.2. Calibration Matrix

In order to determine the calibration matrix C of a device with n qubits, each of the possible $2^n$ states $|s\rangle = |b_0 b_1 \ldots b_{n-1}\rangle$ ($b_i \in \{0,1\}$, i.e. $|s\rangle \in \{0,1\}^n$) must be prepared and measured several times: this results in the distribution of measured values m mentioned above. Figure 19 shows $2^n$ circuits where each circuit $C_s$ prepares the state $|s\rangle$ and measures it: for $|s\rangle = |b_0 b_1 \ldots b_{n-1}\rangle$ the corresponding circuit applies the gate $X^{b_i}$ on qubit i to prepare $|s\rangle$. Immediately after preparing $|s\rangle$, the state is measured.

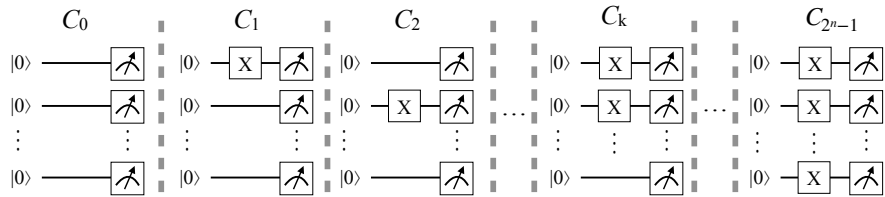

**Fig. 19**. Calibration circuits for determining the calibration matrix (adapted from [46]).

By executing each of the circuits $C_s$ several times the rows ($C_{sj}$) of the calibration matrix C result (see appendix B for concrete examples). Once the calibration matrix C has been determined, its inverse $C^{-1}$ is computed (any method to determine $C^{-1}$ will do).

When a circuit implementing a certain algorithm is executed and its (disturbed) result m has been measured, the formerly derived matrix $C^{-1}$ is applied in a classical postprocessing step and $t = C^{-1} \cdot m$ is returned as the result of the algorithm. I.e. this step corrects the readout error.

The circuits $C_s$ can easily be generated as OpenQASM [13] code, for example. In general, other quantum instruction languages can be used in order to generate the state preparation circuits. This code can be send to and executed on supporting quantum computers to derive the rows $C_s$ of C.

Obviously, the calibration matrix is device dependent. Furthermore, it changes over time (so-called calibration drift). Thus, it must be determined regularly and per device (see Figure 20). The corresponding inverse $C^{-1}$ can be stored and used for postprocessing several times (Figure 21), i.e. there is no need to determine C and compute $C^{-1}$ every time a corresponding postprocessing step is executed.

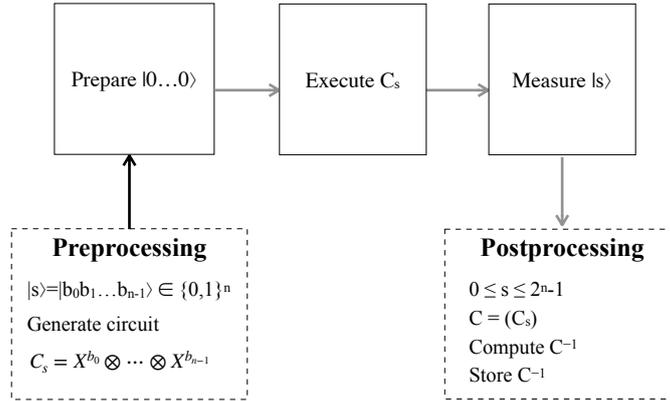

**Fig. 20**. Regularly determining the calibration matrix C and maintaining $C^{-1}$.

## 6.3. Conclusion on Correcting Readout Errors

Readout errors induce disturbances on result distributions of quantum algorithms. Several methods have been proposed to construct a probably undisturbed result distribution from the measured one.

One such unfolding method - the matrix inversion method - can be applied to small NISQ machines: it requires to regularly determine the calibration matrix for a device and invert it, and apply this matrix to every measurement made on this device. Thus, this method imposes both, a regular overhead as well as an overhead for each algorithm executed on the device due to classical postprocessing.

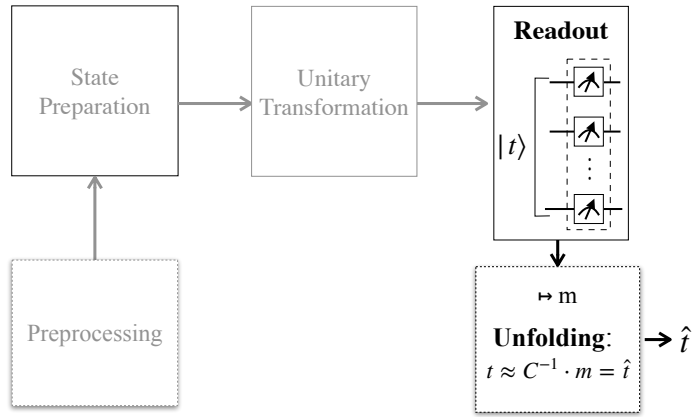

**Fig. 21**. Unfolding the each result read out by means of the retrieved matrix $C^{-1}$.

# 7. Provenance

Provenance subsumes any kind of metadata about artifacts that are relevant for a production process (see [29] for a survey on provenance). The spectrum of what is produced spans from data over workflows to real physical objects. The goal of provenance is to increase understandability, reproducibility, and quality of the end product or its production process [29].

In our context, provenance provides all the information that supports (i) the assessment of the appropriateness of a particular quantum device for successfully executing a circuit, or (ii) the transformation of a quantum algorithm into a circuit (or even a hybrid algorithm) that delivers acceptable results on a particular quantum device (or even in a hybrid environment). For this purpose, lots of information is needed.

First, information about the circuit implementing an algorithm in a hardware-independent manner must be available, like its gates used, its depth and width. Also, the results of the hardware-dependent transpilation process has to be made available. This information is either derived by inspecting the original circuit itself, or is made available on IBM's devices as output of the transpiler.

Second, information about the potential target quantum devices like their number of available qubits is important, their connectivity, the gate set physically supported by a given machine etc. Beyond this static information also dynamic values like the decoherence times of the qubits, the error rates of single qubit operations and two qubit operations, the error rate of the connections between any two qubits etc is key. This information can be directly retrieved from the devices provided by IBM Quantum Experience, for example.

Finally, information subsuming or aggregating, respectively, the processing power of quantum devices is helpful provenance information. For this purpose, different metrics to specify or assess the capability of a quantum computer have been proposed (e.g. quantum volume [14], total quantum factor [61]).

All this information has to be retrieved and made available for transforming or assessing circuits in a uniform manner across vendors of quantum devices and quantum algorithms. Thus, provenance results in a corresponding database containing static information as well as dynamic information that supports the assessment and transformation of circuits.

## 7.1. Conclusion on Provenance

Provenance it needed as a basis for assessing and transforming quantum circuits properly. Part of this information can be determined in a one-time effort like static information about a circuit (e.g. its depth), other parts must be regularly provided like dynamic information about a quantum device (e.g. its calibration matrix).

Figure 22 depicts a simplistic architecture of a corresponding provenance system. An analyzer parses algorithms or circuits, respectively, to derive information about the gates used, depth, width etc. A collector component uses APIs available in quantum environments to retrieve information about the topology of quantum devices, the gates directly supported by the hardware, transpiled circuits etc. An aggregator determines the calibration matrix of a quantum device.

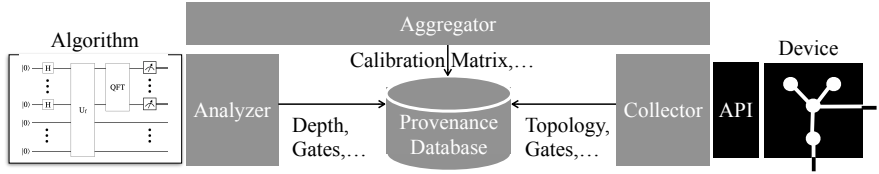

**Fig. 22**. Capturing provenance information.

## 8. Outlook: Error Correction "in the Small"

If errors in a quantum computer would be corrected, most of the work described before would not be necessary. But unfortunately quantum error correction consumes much of the sparse resources available on NISQ machines: additional qubits are required, additional quantum subroutines are needed etc. (see section 8.1). Thus, comprehensive error correction on NISQ machines is unrealistic in practice. However, we discuss below whether error correction "in the small" might increase the quality of results on NISQ machines.

A survey on quantum error correction codes, fault tolerant operations on encoded states, and corresponding circuits is given in [16]. [34] provides a theory of quantum error correction; especially, errors are characterized that can be recovered for a given encoding, and the corresponding recovery operator is constructed. [35] presents a quantum error correction code that allows to detect and recover any error on a single qubit, and this code is minimal (five qubits). A fault tolerant implementation of 1-qubit operations and CNOT as well as fault tolerant measurement is given in [47]. How robustness can be achieved in quantum computations, i.e. arbitrary long computations with any accuracy needed (at the price of a polynomial growing number of qubits), is discussed in [56].

[68] describes a tool that allows to estimate the impact of error correction on algorithms. Performance impact of error correction and its measurement is discussed in [11].

In order to avoid additional qubits (which are rare resources on NISQ devices) for protecting against errors, [19], [72], for example, propose to mitigate errors instead of correcting errors (quantum error mitigation QEM). [7] describe how to use quantum autoencoders to correct the noise of quantum states. They apply their method to various errors of GHZ states.

### 8.1. Error Correction in a Nutshell

In order to detect errors of a physical qubit, several such physical qubits must be bundled into what is called *logical qubit*. The logic qubit will allow to detect and correct errors of the original qubit.

In Figure 23 the qubit q0 contains the state $|\psi\rangle = \alpha|0\rangle+\beta|1\rangle$ to be protected; its state is encoded as $\alpha|000\rangle+\beta|111\rangle$ by means of the encoding subroutine using two additional qubits q1 and q2. Thus, the three physical qubits q0, q1, q2 make up one logical qubit representing $|\psi\rangle$.

Next, the logical qubit is manipulated which takes some time probably resulting in errors in the physical qubits of the logical qubit (note, that the mechanisms discussed here can only protect against errors of a single physical qubit). This is referred to as "noise" in the figure. After that, noise is detected by the syndrome computation circuit, and based on the syndrome appropriate recovery is performed, both collectively called error correction (see e.g. [56] for all the details). In order to avoid modifying the state of the logical qubit by error correction, two additional ancilla qubits are used to hold the error syndrome that steers recovery.

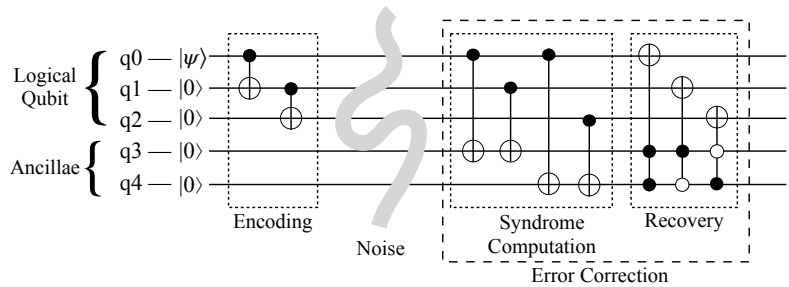

**Fig. 23**. Basic mechanisms of logical qubits and error correction.

Thus, in order to protect a single physical qubit against errors, four additional qubits are needed as well as additional gates that perform the proper encoding and correction (see [16] for an extensive treatment of the subject). Furthermore, the gates of the proper algorithm must now act on logical qubits instead of physical qubits which means that these gates must be realized by corresponding subroutines. On the left side of Figures 24 a circuit with a 1-qubit gate G1 and a 2-qubit gate G2 is shown. Each manipulated physical qubit is turned by encoding into a logical qubit requiring five qubits each (including ancillae qubits). As a result, gates G1 and G2 must be realized by subroutines S1 and S2 each of which is realized by a separate circuit. After executing such a subroutine, error correction (EC) on the logical qubits must be performed. The resulting error corrected circuit is depicted on the right of the figure.

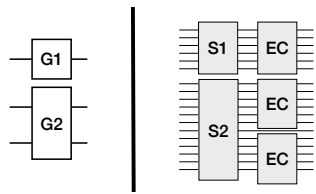

**Fig. 24**. Error correction implies an increase in depth and with.

### 8.2. Protecting Individual Qubits

Thus, dealing with logical qubits implies a lot of overhead in terms of both, depth and width of the resulting algorithm which is too expensive for todays NISQ machines (remember from section 1.1: width times depths must be very small).

As an (admittedly ad hoc) hypothesis, it might suffice to turn only a subset of the physical qubits of a circuit into logical qubits, i.e. only individual "important" physical qubits might be protected by error encoding and error correction. This hypothesis is suggested by the fact that a couple of algorithms like Quantum Phase Estimation perform most of their operations on a few qubits only, i.e. these few qubits seem to be the important ones.

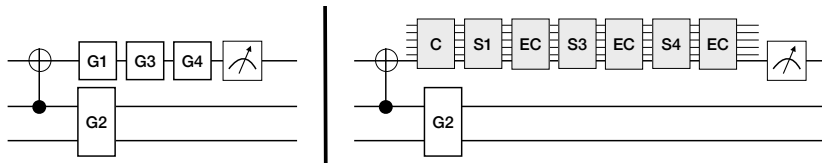

**Fig. 25**. Error correction on individual qubits only.

Figure 25 depicts the mechanism of protecting individual qubits: it gives a schematic algorithm on the left side the first qubit of which seems to be important and should be protected. The right side of the figure shows the resulting circuit: the qubit that is deemed important is encoded by the encoding subroutine C after the CNOT. Gates G1, G3 and G4 acting on the first (i.e. the important) qubit are substituted by the corresponding subroutines S1, S3 and S4. After each subroutine an error correction subroutine is injected. Finally, one of the qubits bundled into the logical qubit (i.e. not one of the ancillae) is measured.

Appendix C discusses experiments we run on quantum computers to check whether error correcting individual qubits can be realistically performed. It turned out from these experiments that on small machines this is unrealistic being too erroneous. Consequently, we were not able to prove or disprove whether our ad hoc hypothesis is correct.

### 8.3. Conclusion on Error Correction

Full error correction cannot applied on todays NISQ machines because of their lack of quantum resources. Even protecting individual (e.g. important) qubits only increases the error of already simple algorithms revealing even the restricted variant of error correction "in the small" as impractical.

## 9. Conclusion and Future Work

We emphasized the significance of several aspects often not in the foreground when discussing quantum algorithms and their potential execution on todays NISQ machines. Preparing the input to be processed by a quantum algorithm may require significant classical preprocessing to generate the circuit that is used for state preparation on the quantum computer. This circuit may significantly increase the depth of the circuit implementing the quantum algorithm proper.

Quantum algorithms that use subroutines (especially oracles) need these subroutines to be specified as a circuit. As before, these circuits may contribute crucially to both, the depth and width of the algorithm proper.

The topology graph of the target device of a circuit has increasing effects on the depth of an algorithm too. Even worse, this impact is time-dependent.

There is potential to optimize an algorithm in a hardware-independent manner by reducing its depth and width. A simple method like "shifting gates to the left" is already helpful. Also, the transpiler of the target device has impact on an algorithm's depth and should be considered.

Reading out the result produced by an algorithm is erroneous too. Improving the precision of such erroneous results requires to run regular quantum circuits to determine the calibration matrix of a target device, and to apply this calibration matrix to every result produced on this device. The latter is done in a classical postprocessing step. Note, that there is no impact on the quantum algorithm proper by correcting readout errors via the presented matrix inversion method, but the quantum device is additionally loaded by determining this matrix (likely imposing cost).

Most of these tasks require metadata: we briefly sketched provenance and how this metadata can be derived.

Finally, we hypothesized about error correction of individual qubits: since full error correction is out of scope for NISQ devices this might turn out to be beneficial. This will increase the depth and width of the corresponding algorithm, and even simple error correction turned out to be impractical.

We described several experiments that we performed on concrete quantum computers to underpin some of our arguments.

Currently, we are developing a pattern language for quantum computing [37] in which we fold in most of the results presented here. These patterns, but also application prototypes using these patterns will be published on a corresponding platform [38] that is developed in course of the PlanQK project [55]. Furthermore, the methods discussed are part of a so-called NISQ Analyzer under construction that will assess whether a given algorithm may execute successfully on a certain quantum computer. A tool for capturing provenance information is under construction too.

**Acknowledgements**

We are very grateful to our colleagues Marie Salm, Benjamin Weder, Manuela Weigold, and Karoline Wild for discussing with us selective subjects of this paper. Also, we are grateful to IBM for providing open access to its quantum computers.

This work was partially funded by the BMWi project PlanQK (01MK20005N).

## Appendix A: Impact of Error Rates on IBM Quantum Experience

In order to show the impact of a topology of a particular quantum computer on the rewrite of a circuit and initial qubit allocation (see sections 4.2 and 4.3), we performed a simple experiment on three devices of the IBM Quantum Experience. First, to see the effect of different success rates of the connections on qubit movement and initial qubit allocation we apply a CNOT to two qubits that are not adjacent in the topology graph. Second, the influence of the success rate of single qubit operations on qubit allocation should be shown. Third, the time dependency of a topology, i.e. the modifications of the success rates in the topology (e.g. by calibration), and its impact on the circuit rewrite should be seen.

Figure 26 depicts the simple circuit of our experiment: qubit q[1] is negated by a Pauli X gate, next a controlled-not CNOT(q[1], q[4]) is applied with control qubit q[1] and target qubit q[4]. Finally, q[4] is measured in the computational basis (the measurement is irrelevant for what follows).

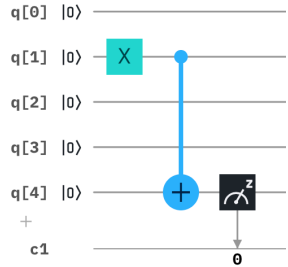

**Fig. 26**. Circuit used to show the impact of the topology graph on circuit rewrite.

### A.1  ibmq_oursense

In our first experiment we used the device ibmq_ourense (with 5 qubits) the topology graph of which is shown on the left side of Figure 27 to compile our circuit. The success rates of CNOTs and the 1-qubit U2 gate (which is directly supported by the hardware) are indicated by a color code. The transpilation result is on the right.

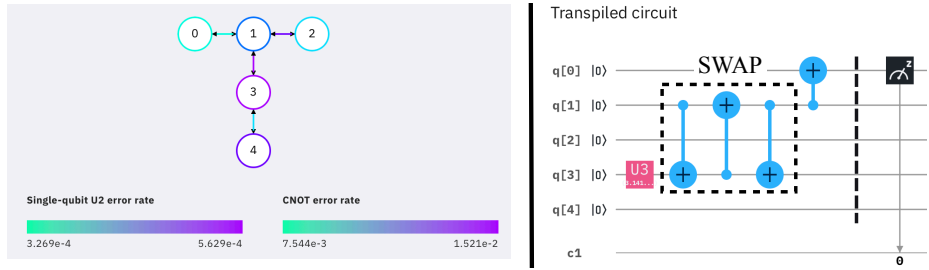

**Fig. 27**. Topology graph of ibmq_ourense and the transpilation result.

The following (simplified) reasoning helps to understand the result of the transpilation. Transpilation strives towards minimizing errors of 1-qubit gates and CNOTs. Because CNOT error rates are higher than error rates of 1-qubit gates, qubits are allocated to positions where errors of required CNOTs are minimized, sometime choosing qubit positions with higher error rates of 1-qubit gates although qubits with less error rates of 1-qubit gates are available in the topology.

Hence, the circuit variable q[1] is allocated to the physical qubit *3* of the machine (although qubits with less error rates like qubit *2* are available), and q[4] is allocated to the physical qubit *0*. Effectively, q[1] ↦ *3* and q[4] ↦ *0*. As the color code of the topology graph in the figure indicates, qubit *3* has a good (not the best) success rate for the two-rotation operation that is directly supported by the devices of the IBM Quantum Experience. Thus, this qubit is chosen for applying the X gate there.

Note that the Pauli X gate is performed in the transpiled circuit as an $U3(\pi,0,\pi)=X$ on *3* (i.e. as a two-rotation gate like U2, thus, the success rate of U2 applies), where

$$U3\left(\theta, \varphi, \lambda\right) = \begin{pmatrix} \cos\frac{\theta}{2} & -e^{i\lambda}\sin\frac{\theta}{2} \\ e^{i\varphi}\sin\frac{\theta}{2} & e^{i(\lambda+\varphi)}\cos\frac{\theta}{2} \end{pmatrix}$$

Furthermore, the connection between qubits *0* and *1* has very high success rate in the topology, thus, the CNOT should be applied here. For this purpose, qubit *3* (=q[1]) is swapped to qubit *1* (effectively turning qubit *1* into q[1]: *1*=q[1]), thus, CNOT(*1*,*0*) ≜ CNOT(q[1],q[4]). Recall that a SWAP is a sequence of three CNOTs (section 4.1), i.e. the SWAP is represented in the rewritten circuit accordingly.

### A.2   ibmq_london

The next experiment was performed on the device ibmq_london, which has also 5 qubits and the structure of the topology graph is the same as the one of ibmq_oursense. The topology graph is shown on the left side of Figure 28 indicating the success rates. As before, the transpilation result is on the right.

As before, the initial allocation of qubits assigned q[1] to the physical qubit *3*, and q[4] to the physical qubit *0*, i.e. q[1] ↦ *3* and q[4] ↦ *0*. Qubit *3* has a good enough success rate for the two-rotation operation, thus, this qubit is chosen for applying the U3($\pi$,0,$\pi$)=X gate there. In this case, the connection between qubits *1* and *3* has very high success rate, thus, the CNOT should be applied here. For this purpose, qubit *0* (=q[4]) is swapped to qubit *1* (i.e. *1*=q[4] afterwards), thus, CNOT(*3*,*1*) ≜ CNOT(q[1],q[4]).

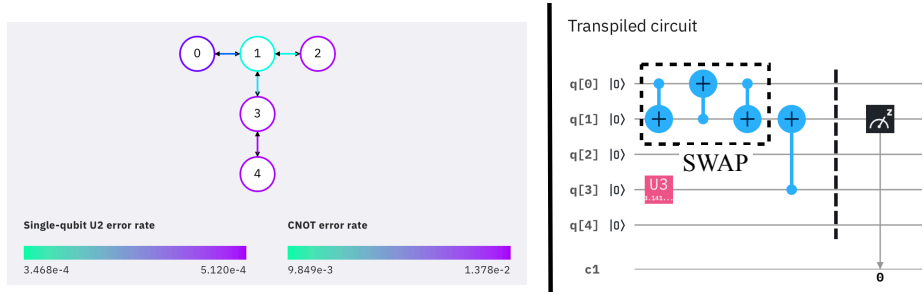

**Fig. 28**. Topology graph of ibmq_london and the transpilation result.

### A.3   ibmq_london (After Calibration)

The next experiment elucidates the time dependency of the various success rates and the resulting time dependency of transpilation. Again, we used the device ibmq_london once it got calibrated after the last experiment. The weights and colors of the topology graph changed and is shown as before on the left side of Figure 29 , and the transpilation result is on the right side of the figure.

The initial allocation of qubits is the same (i.e. q[1] ↦ *3* and q[4] ↦ *0*): qubit *3* has a good success rate for U3($\pi$,0,$\pi$)=X, and qubit *0* is used because connection (*0*,*1*) is selected for the second step, allocating q[4] to *0* avoids another SWAP. After recalibration, the connection between qubits *0* and *1* has very high success rate making it the target for applying CNOT. Thus, qubit *3* (=q[1]) is swapped to qubit *1* (i.e. *1*=q[1] afterwards), thus, CNOT(*1*,*0*) ≜ CNOT(q[1],q[4]).

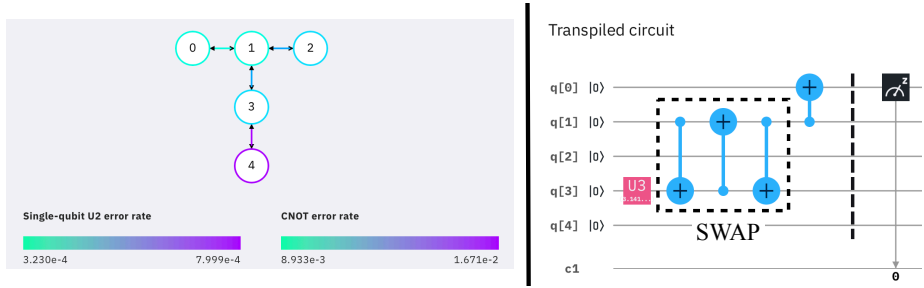

**Fig. 29**. Topology graph of ibmq_london and the transpilation result after recalibration.

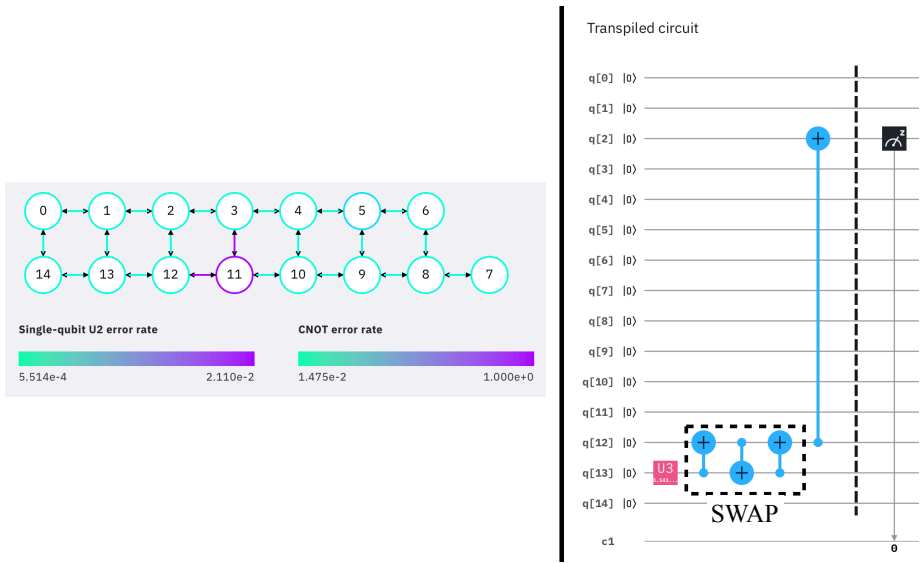

**Fig. 30**. Topology graph of ibmq_16_meldbourne and the transpilation result.

### A.4   ibmq_16_melbourne

Finally, we wanted to explore the topology sensitivity of a device with a very different topology and used ibmq_16_melbourne, which is a 15 qubit device. Its topology graph is shown on the left side of Figure 30, and the transpilation result is on the right side of the figure.

The initial allocation of qubits assigned q[1] to the physical qubit *13*, and q[4] to the physical qubit *2*, i.e. q[1] ↦ *13* and q[4] ↦ *2*. Qubit *13* has a high success rate for applying U3($\pi$,0,$\pi$)=X (especially higher than qubit *12*). The connection between qubits *12* and *2* has very high success rate, thus, the CNOT is applied here. For this purpose, qubit *13* (=q[1]) is swapped to qubit *12* (i.e. *12*=q[1] afterwards), thus, CNOT(*12*,*2*) ≙ CNOT(q[1],q[4]).

## Appendix B: Measuring the Calibration Matrix

We measured selective rows of the calibration matrix of the ibmq_rome device. For this purpose we used the circuit $C_s$ (see section 6.2), executed it on the device with 1024 shots, and retrieved the histogram of the result distribution. The histogram depicts percentages for values that occurred above a threshold, i.e. values that have been measured less than this threshold are considered to have not been measured at all. The values measured above this threshold are denoted on the x-axis of the distribution, and the size of the corresponding bins correspond to the percentage their values occurred. Circuit $C_s$ prepared the state |s⟩, i.e. it represents the value s, and if the value j has been measures instead of s in x% of all shots, then the matrix element $C_{sj}$ is set to x. This way, circuit $C_s$ produces the s-th row of the calibration matrix C.

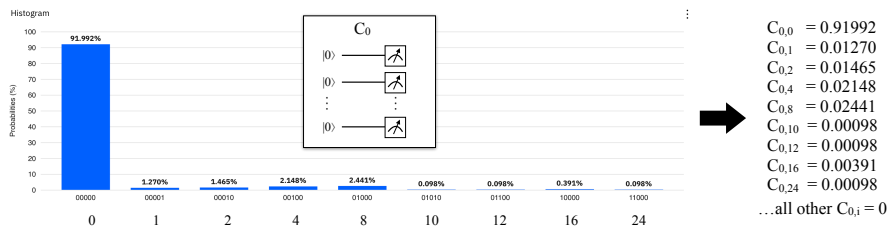

**Fig. 31**. Measuring the C0-row of the calibration matrix of ibmq_rome.

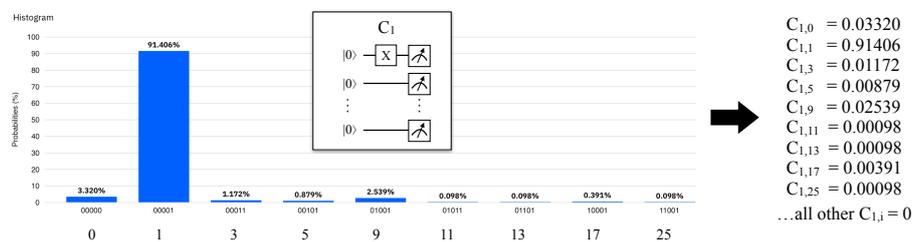

**Fig. 32**. Measuring the C1-row of the calibration matrix of ibmq_rome.

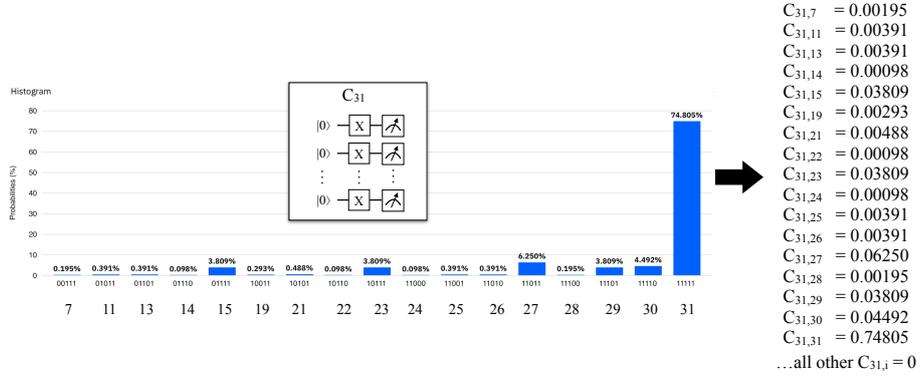

**Fig. 33**. Measuring the C31-row of the calibration matrix of ibmq_rome.

## Appendix C: Error Correction "in the Small"

The following experiments have been performed to check whether or not individual qubits are worth to be protected against errors on NISQ machines. For this purpose we first implemented the error correction circuit discussed in section 8.1 in the circuit composer of IBM Quantum Experience (see Figure 34). The depth of the circuit shown is 14. The circuit prepares the state $|1\rangle$ for qubit q[0] and then encodes it with the three qubit code. Next the syndrome is computed and the recovery takes place. When performing this circuit on the simulator it produces the correct result, i.e. the bit string 00111 is measured with 100% probability.

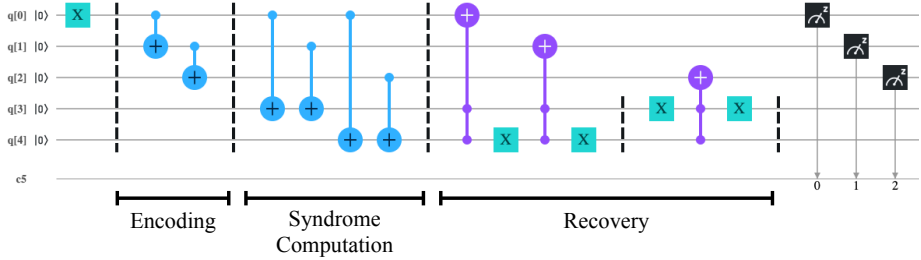

**Fig. 34**. Error correction circuit.

We introduced noise between encoding and syndrome computation by flipping the qubit q[1] (see Figure 35). Again, the simulator delivered the correct result 00111 with 100% probability, i.e. the bit flip error was correctly detected and recovered.

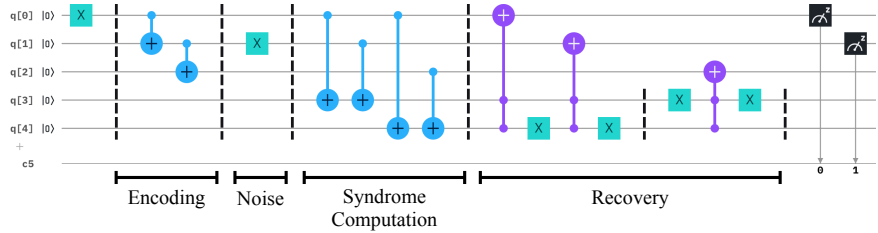

**Fig. 35**. Error correction circuit correcting the noise.

Next, the error correction circuit was transpiled for the ibmq_16_melbourne device. Figure 36 shows the result: the figure is not intended to be read but to give an impression only. The depth of the resulting transpiled circuit is 76.

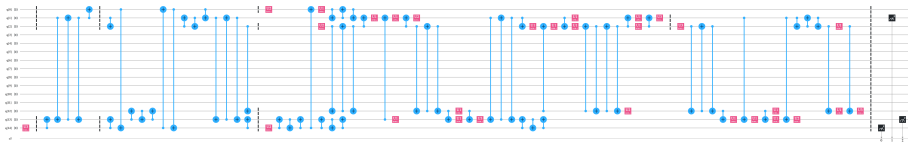

**Fig. 36**. Result of transpiling the error correction circuit on ibmq_16_melbourne.

The result from executing the transpiled circuit on ibmq_16_melbourne with 8192 shots was not encouraging: the probability of the correct result 00111 was only 10.229% (Figure 37). Several other (erroneous) results occurred with higher probability.

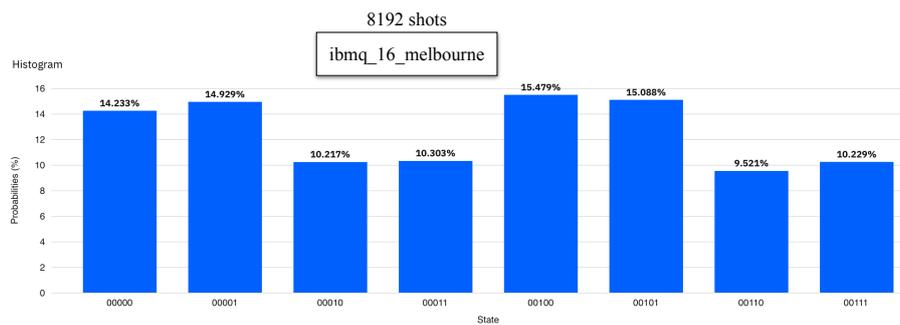

**Fig. 37**. Result executing the error correction circuit without noise on ibmq_16_melbourne.

Similarly, executing the circuit with noise on ibmq_16_melbourne with 8192 shots was not encouraging either: the probability of the correct result 00111 was only 7.312% (Figure 38). As before, several other (erroneous) results occurred with higher probability.

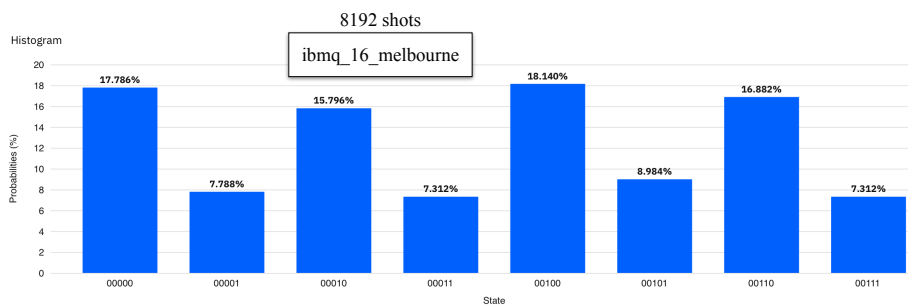

**Fig. 38**. Result executing the error correction circuit with noise on ibmq_16_melbourne.

Thus, our simple experiments do not suggest to consider error correction for protecting even individual qubits on NISQ machines. One reason for this might be the significant increase in depth caused by the error correction circuit that has to be injected into the circuit the qubit of which should be protected.